\documentclass[
reprint,
runinaddress,
preprintnumbers,
nofootinbib,
 amsmath,amssymb,amsthm,
 aps,
prd,
]{revtex4-1}
\usepackage{lipsum}
\usepackage{dcolumn}
\usepackage{bm}
\usepackage{tikz}
\usetikzlibrary{matrix}
\usepackage{slashed}
\usepackage{fontenc}
\usepackage{graphicx,hyperref,color,xcolor}
\usepackage[scr=rsfso,cal=zapfc,frak=euler,bb=ams]{mathalfa}

\definecolor{awesome}{rgb}{1.0, 0.13, 0.32}
\definecolor{electricblue}{rgb}{0.03, 0.57, 0.82}
\definecolor{guppiegreen}{rgb}{0.0, 0.88, 0.4}
\definecolor{blue-violet}{rgb}{0.54, 0.17, 0.89}

\makeatletter
\def\p@figure{\color{awesome}}
\def\p@equation{\color{electricblue}}
\makeatother

\hypersetup{
    bookmarks=true,         
    unicode=false,          
    pdftoolbar=true,        
    pdfmenubar=true,        
    pdffitwindow=false,     
    pdfstartview={FitH},    
    pdfnewwindow=true,      
    colorlinks=true,       
    linkcolor=blue-violet,          
    citecolor=guppiegreen,        
    filecolor=blue,      
    urlcolor=guppiegreen          
}

\newcommand{\be}{\begin{equation}}
\newcommand{\ee}{\end{equation}}
\newcommand{\ba}{\begin{eqnarray}}
\newcommand{\ea}{\end{eqnarray}}
\newcommand{\basplit}{\begin{eqnarray}\begin{split}}
\newcommand{\easplit}{\end{split}\end{eqnarray}}
\newcommand{\bes}{\begin{equation*}}
\newcommand{\ees}{\end{equation*}}
\newcommand{\beas}{\begin{eqnarray*}}
\newcommand{\eeas}{\end{eqnarray*}}
\newcommand{\bas}{\begin{array*}}
\newcommand{\eas}{\end{array*}}

\usepackage{titlesec}
\titleformat*{\paragraph}{\small\bfseries\itshape}

\begin{document}

\preprint{YITP-21-90}
\title{{\large{Information storage and near horizon quantum correlations}}}
\author{Abram Akal}
\email[Current e-mail: ]{a.akal@uu.nl}
\thanks{Current affiliation: Institute for Theoretical Physics, Utrecht University, Princetonplein 5, 3584 CC Utrecht, The Netherlands}
\affiliation{Center for Gravitational Physics\\
Yukawa Institute for Theoretical Physics\\
Kyoto University, Kyoto 606-8502, Japan\\}
\date{\today}

\begin{abstract}
It is usually stated that the information storing region associated with the Bekenstein-Hawking entropy is enclosed by a sphere of diameter equal twice the gravitational radius. We point out that this cannot apply to a quantum black hole. The deviation is particularly revealed when the latter is maximally correlated with its Hawking radiation. Specifically, we demonstrate that the size of the entropy sphere associated with the underlying microstructure has to be necessarily broadened when the fine grained radiation entropy becomes maximal. Such an enlargement is understood to be the consequence of unitarization effects in quantum gravity and aligns with recent findings in holography arguing that purification happens via semiclassically invisible quantum correlations extending across the black hole atmosphere. In the present work, we consider an evaporating black hole in asymptotically flat spacetime. We assume that the standard thermodynamical description is valid so long the black hole viewed from the outside is sufficiently large, radiation escaping into the future null infinity can be described on a smooth spacetime background, and the von Neumann entropy of Hawking radiation evolves unitarily. We briefly comment on the black hole singularity.
\end{abstract}

\maketitle


\section{Introduction}
\label{sec:intro}

Even though quantum gravity is still not sufficiently well understood, there has been significant progress during the past decades suggesting a holographic formulation. Roughly speaking, the latter proposes that physics associated with some spacetime region can be formulated in terms of the degrees of freedom living on the boundary of that region \cite{Bekenstein:1973ur,Bekenstein:1980jp,tHooft:1993dmi, Susskind:1994vu}. Gravitational holography \cite{Bousso:1999cb,Bousso:2002ju} has found its explicit realization within string theory in form of the AdS/CFT correspondence \cite{Ma,Gubser:1998bc,Witten:1998qj} arguing that gravity in asymptotically anti-de Sitter (AdS) space has a dual formulation in terms of a single conformal field theory (CFT) defined on the AdS boundary.

The formulation of quantum gravity is often believed to be relevant for understanding physics at very high energies or in the vicinity of spacetime singularities. However, this expectation may fall too short from various perspectives. It seems more adequate to accept that successfully describing gravity using the principles of quantum mechanics will not only teach us about the extreme limits that spacetime geometry is exposed to. Instead, a proper formulation of quantum gravitational physics should also incorporate knowledge about certain quantum correlations that will be relevant in situations far away from the described instances. This goes back to a nontrivial mixing of energy scales and is, at least, what seems to be enforced by holography. Such an understanding resonates with more recent developments relying on the holographic entanglement entropy proposal \cite{RT} and its generalizations \cite{Hubeny:2007xt,Faulkner:2013ana}.

In the present work, starting from the discussion in \cite{Akal:2020ujg}, we further advocate that precisely such kind of quantum gravitational effects turn out to be highly relevant in understanding the nature of quantum black holes.

The classical treatment of black holes led to tensions between the principles of quantum mechanics and general relativity.
Bekenstein laid the ground \cite{Bekenstein:1972tm,Bekenstein:1973ur} for the laws of black hole mechanics \cite{Bardeen:1973gs} and soon after, Hawking semiclassically showed that black holes release thermal radiation, and eventually completely evaporate if the total energy is conserved \cite{Hawking:1974rv,Hawking:1974sw}. These insights gave rise to the Bekenstein-Hawking formula\footnote{Here, $A_\text{h}$: black hole's horizon surface area; $k$: Boltzmann constant; $c$: speed of light; $G$: Newton constant; $\hbar$: reduced Planck constant. In the remaining part, we set $k = c = 1$.} 
\ba
S_\text{BH} = \frac{A_\text{h} k c^3}{4 G \hbar}
\label{eq:S_BH}
\ea
computing the coarse grained entropy $S_\text{BH}$ of a black hole of finite mass \cite{Bekenstein:1973ur,Hawking:1974sw}.
However, since, in principle, a black hole can be formed from a pure state, arguing that radiation ends up being thermal after evaporation initiated a severe inconsistency with unitarity in quantum mechanics.\footnote{Within the semiclassical approach, respecting the laws of black hole mechanics, energy conservation will be maintained. However, what the semiclassical computation does not account for, is the correct amount and structure of quantum information encoded in the radiation that is required for unitarity.} 

Such findings have led to many speculations about whether or not unitarity is maintained in gravitational physics \cite{Hawking:1976ra}. This puzzle is often referred to as the black hole information paradox.

An argument which may indicate that the evaporation of a black hole formed from a pure state is a unitary process, has been proposed by Page \cite{Page:1993df,Page:1993wv}. 
The starting point is the idea that \eqref{eq:S_BH} is expected to give the leading term for the logarithm of the number of microstates associated with the black hole. Accordingly, \eqref{eq:S_BH} should be taken as the amount of entanglement entropy that the black hole can maximally possess. When the number of quantum states of Hawking radiation becomes larger than the number of the remaining black hole states, which is expected to happen after the so-called Page time $u_\text{P}$, the von Neumann entropy of the radiation, $S_\text{rad}$, should be limited by the entropy of the evaporating black hole and, in fact, would approach it from below, $S_\text{rad} \leq S_\text{BH}$.
The curve which describes the evolution of the fine grained radiation entropy, and approaches zero in the past and future time limits, i.e. $S_\text{rad} = 0$, is known as the Page curve. The increasing-decreasing property of the entropy curve does not only apply if the initial matter is pure. A similar behavior applies to an evaporating black hole that has formed from a mixed state, where $S_\text{rad}$ again starts from zero but later saturates at the non-zero entropy associated with the initial state \cite{Page:2013dx}.\footnote{In fact, starting from a generic black hole metric cannot distinguish between these two scenarios. Knowledge about the history of formation is necessary and would require the S matrix in a full quantum gravity description.}

During the evaporation process the horizon surface area decreases since the expectation value of the matter stress energy tensor, $\langle T_{ab} \rangle$, may violate the (semiclassical) null energy condition (NEC),
$\langle T_{ab} \rangle k^a k^b \geq 0$ with $k^a$ being some null vector, namely, even in cases when the NEC holds classically. This contradicts with the main hypothesis of the area theorem $dA_\text{h} \geq 0$, which would mean a failure of the second law of black hole mechanics. However, the notion of a generalized entropy $S_\text{gen}$ introduced by Bekenstein \cite{Bekenstein:1973ur}, which is nothing but the gravitational black hole entropy plus the matter (von Neumann) entropy $S_\text{out}$ outside the horizon,
\ba
S_\text{gen} = S_\text{BH} + S_\text{out},
\label{eq:S_gen}
\ea
evades such a violation. Although $A_\text{h}$ will decrease, this cannot compensate the radiation generated increase in the von Neumann entropy outside, i.e. $dS_\text{out} \geq -dS_\text{BH}$, \cite{Page:2013dx}. Hence, even though the standard second law fails in the presence of classical black holes, as well as the second law of black hole mechanics if quantum effects are considered, the generalized second law (GSL) \cite{Bekenstein:1974ax}, which is the statement that
\ba
dS_\text{gen} \geq 0,
\label{eq:GSL}
\ea
may hold in the semiclassical limit \cite{Sorkin:1986mg,Wall:2011hj}. 

\begin{figure}[h]
  \centering
    \includegraphics[width=0.18\textwidth]{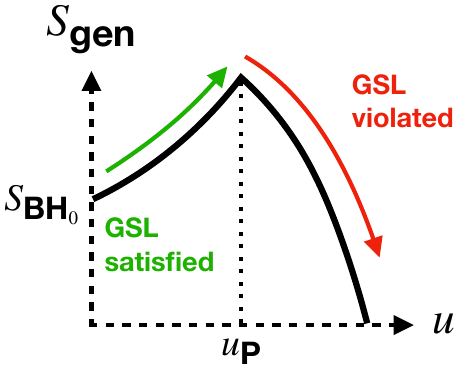}
    \caption{Sketch of time ($u$) evolution of the generalized entropy $S_\text{gen}$ in the case of a unitarily evaporating black hole. While the GSL will be satisfied for $u \leq u_\text{P}$ (green), it will be violated for $u > u_\text{P}$ (red). Here, $u_\text{P}$ corresponds to the Page time.}
    \label{fig:Sgen}
\end{figure}

Despite the fact that both $S_\text{BH}$ and $S_\text{out}$ are cutoff dependent separately, \eqref{eq:S_gen} itself is finite, because the divergences cancel out \cite{Susskind:1994sm,Jacobson:1994iw,Fursaev:1994ea,Demers:1995dq,Kabat:1995eq,Larsen:1995ax}. 
This suggests an important role of $S_\text{gen}$ in a full quantum gravity description, where the gravitational part may account for Planckian degrees of freedom \cite{Frolov:1993ym,Susskind:1994sm,Jacobson:1994iw}. 
A unitary evaporation process consistent with Page's arguments would necessarily violate the GSL \eqref{eq:GSL}.\footnote{Again, if $S_\text{out}$ in \eqref{eq:S_gen} is taken to be the fine grained entropy of the radiation.} Hence, there has to exist a mechanism going beyond the semiclassical description that explains how the radiation gets purified. In fact, to make \eqref{eq:S_gen} finite and continuously decreasing for an old black hole as shown in Fig.~\ref{fig:Sgen}, this should be required.\footnote{Semiclassically, $A_\text{h}$ has no special feature. 
From the perspective of an observer falling into the black hole, there is no specialty about the horizon. However, unitarily, seen from the distance, or more precisely, from the holographic perspective, its location necessitates more quantum information structure.
Associating some coarse grained entropy with it cannot be sufficient in addressing the unitarity puzzle. Particularly,
one has to understand the connection between the underlying degrees of freedom and the thermal nature of Hawking radiation. As we will point out, such entities may have an intricate entanglement structure and give rise to locally inaccessible quantum correlations that cannot be captured in a semiclassical description. In fact, integrating out this near horizon quantum information leads to thermalization and thus explains the state mixture seen in the semiclassical computation \cite{Akal:2020ujg}. Refer also to \cite{Nomura:2019qps,Nomura:2020ska} discussing the role of the atmosphere region in the thermalization process.

On the other hand, it is the (apparent) horizon, which, by definition, is the place behind where events stop being observable \cite{Rindler:1956yx}. The former defines the gravitational part in the generalized entropy. It is the physical surface determining the evaporation rate of the black hole respecting the standard laws of thermodynamics, at least so long the black hole is large enough, which would surely be the case around $u_\text{P}$.}

\begin{figure}[h]
  \centering
    \includegraphics[width=0.35\textwidth]{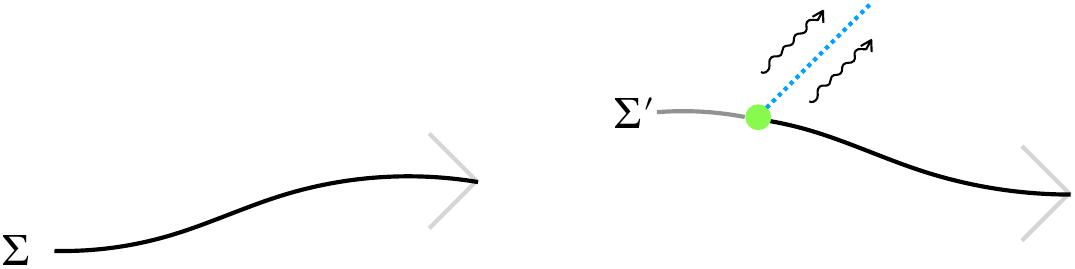}
    \caption{Semiclassical gravity: Before the black hole has formed, any Cauchy surface $\Sigma$ (left) describing the initial pure state can be evolved unitarily. After the black hole has formed by gravitational collapse, radiation will be continuously generated (right). Correlations between the exterior and the interior are built up only in the form of quantum entanglement between the outgoing Hawking particles and their interior partners causally disconnected by the black hole horizon (green dot). After the evaporation process, any constant time slice $\Sigma''$ cannot be completed due to the Cauchy horizon splitting the latter into two parts, cf. Fig~\ref{fig:evapBHsemi}. Any such constant time slice $\Sigma''$ cannot be traced back to a Cauchy surface $\Sigma'$ in the pre-Planckian phase of the black hole due to the existence of the singularity. An initial pure state will evolve into a mixed state.}
    \label{fig:SplitSemi}
\end{figure}

Classically, matter forming the black hole ends up in the black hole singularity. Semiclassically, all interior-exterior correlations are generated while the black hole radiates, see Fig.~\ref{fig:SplitSemi}. This entanglement structure leads to thermal radiation and does not capture the information content required for unitarity. The evolution of a pure state to a final mixed state in the formation and evaporation process may be understood as the result of a final time slice that cannot be taken as a Cauchy surface due to the singularity.
However, at least when the GSL \eqref{eq:GSL} becomes violated, as shown in Fig.~\ref{fig:Sgen}, there would be no reason to expect the notion of a singularity lying within the deep interior of the black hole. The violation occurs much before the Planckian final stage of the black hole. Accordingly, the information structure associated with the (holographic) global pure state should become visible to the outgoing Hawking particles.
This suggests that, when the black hole is still sufficiently massive, assuming that the initial mass is large enough, the notion of a classical singularity will already be ill-defined. In this sense, it is unclear whether, and if yes, how imposing a final state boundary condition on the singularity as proposed in \cite{Horowitz:2003he} can be implemented.

We think that the mechanism responsible for the purification has to play an important role in the singularity resolution as well, assuming that the radiation entropy evolves as described. Indeed, such an evolution seems to be inevitable, at least, when \eqref{eq:S_BH} is associated with some microstructure.\footnote{Namely, in order to arrive at Page's argument, one has to accept that $S_\text{BH}$ accounts for the number of microstates associated with the black hole, which is a direct implication of the holographic principle. This, in fact, is the reason why the entanglement entropy of black hole radiation is expected to be bounded from above by $S_\text{BH}$ at $u=u_\text{P}$. The assigning of a microstructure quantified by $S_\text{BH}$ is the very essence behind the Page curve. Otherwise, there would be no reason to expect such an evolution curve in a unitarity description. The finiteness of the sum of the gravitational term $S_\text{BH}$ and the radiation entanglement entropy $S_\text{rad}$ has extensively been discussed, see e.g. \cite{Susskind:1994sm}, supported by microscopic derivations of $S_\text{BH}$, starting with the early progress in \cite{Strominger:1996sh}.

We shall emphasize that in the particular case of the AdS/CFT correspondence, $S_\text{BH}$, i.e. the gravitational term appearing in Bekenstein's generalized entropy, \eqref{eq:S_gen}, may determine the number of superselection sectors in the dual boundary CFT under the action of a restricted (semiclassical) subalgebra \cite{Akal:2020ujg}, and does therefore align with the definition of topological entropy $S_\text{topo}$ as, for instance, discussed in \cite{Kitaev:2005dm}. Accordingly, the combination of both entropies, i.e.
$S_\text{BH} + S_\text{rad}$, may be expected to be a finite, fine-grained quantity that captures the information contained in the complete unitary description. Notably, in a topologically ordered quantum system, the fine grained entanglement entropy of a subregion can be viewed as the usual area term plus the topological term $S_\text{topo}$. 

Of course, one might refuse the assumption of an underlying microstructure, but then there would be no point in discussing the importance of the Page curve. As one of the basic assumptions in the present work, we assume that $S_\text{BH}$ quantifies the underlying (topologically) protected microscopic information, and, thus, validates the notion of a Page curve in the process of black hole evaporation.} 

\begin{figure}[h]
  \centering
    \includegraphics[width=0.35\textwidth]{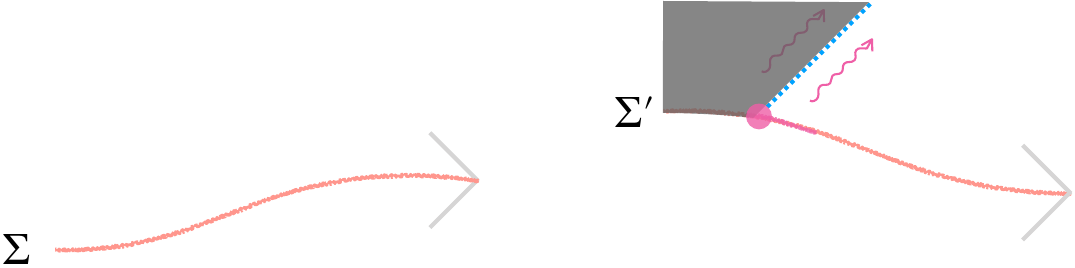}
    \caption{Gravitational holography: Before the formation of the horizon, any Cauchy surface $\Sigma$ represents a pure quantum state (left). This purity on the gravitational side shall be holographically formulated in terms of a quantum mechanical system. One may assign a complete algebra to the dual quantum theory. A semiclassical spacetime description, see Fig.~\ref{fig:SplitSemi}, emerges as a result of reducing the complete algebra to a subalgebra that does not maintain bulk unitarity. However, in every instance, the gravitational side holographically possesses a full quantum description preserving unitarity. Especially, the stage after the evaporation can be traced back unitarily to any previous stage at which the evaporation is still ongoing. In order to allow for this unitary evolution, there will exist certain quantum correlations that cannot be captured by the subalgebra. Utilizing the holographic entanglement entropy proposal and related measures, it has been pointed out that such semiclassically invisible correlations should reside across the black hole atmosphere \citep{Akal:2020ujg}. As schematically shown above, it is this near horizon information structure, getting encoded in the outgoing Hawking particles (right), which gives rise to a unitary entropy curve. The shading in the right panel shall illustrate that the notion of an interior region does not apply under the action of the complete algebra.}
    \label{fig:SplitHolo}
\end{figure}

Within AdS/CFT, we have pointed out that unitarization during evaporation proceeds via semiclassically invisible quantum correlations extending across the black hole atmosphere \cite{Akal:2020ujg}. The existence of such quantum correlations would go back to the presence of certain, necessarily extended horizon entities that will not belong to the algebra describing semiclassical bulk physics. 
Based on findings relying on the (generalized) holographic entanglement entropy proposal and related measures, it has been discussed that these atmosphere correlations give rise to a unitary entropy curve. 
Accordingly, the violation of the GSL would be induced due to the transfer of this near horizon information structure to the escaping low energy Hawking particles, see Fig.~\ref{fig:SplitHolo}.
The presence of Planckian horizon physics resonates with earlier considerations in string theory \cite{Susskind:1993ws,Susskind:1994sm,Horowitz:1996nw} which indeed indicated that black holes should be viewed as ordinary quantum systems.

In this work, we provide further evidence supporting the purification mechanism described above. The strategy is the following. If there indeed exist certain near horizon quantum correlations responsible for the unitarization process, then, following the discussion in \cite{Akal:2020ujg}, the system that will be correlated with the radiation should start appearing maximally enlarged when $u = u_\text{P}$, thus, exceeding the size of the thermodynamic sphere.\footnote{The thermodynamic sphere is defined through the black hole's gravitational radius.} 

We show that, if the total energy is conserved and unitarity, means, the right amount and structure of quantum information is encoded in the outgoing radiation, is maintained, such an understanding follows in an exterior spacetime picture. We do not fix any specific details of the setup, so that the conclusions presented here are expected to have universal validity. 

\section{Entropy, energy, spheres}
\label{sec:ees}

To start with, suppose that a spherical uncorrelated matter system with radius $R$ and energy $E$ is going to be added to a stationary Schwarzschild black hole. The system shall be carefully lowered slowly towards the horizon before being dumped in. In this way, the system's mass may effectively be lowered due to gravitational redshifting by which, if this is possible all the way down to the horizon, $E$ may be extracted as work. Finally, dropping it in, would not increase the black hole mass and, thus, violate the GSL \eqref{eq:GSL}.
In order to have the latter satisfied, Bekenstein argued that
the entropy-to-energy ratio should be bounded as follows \cite{Bekenstein:1974ax,Bekenstein:1980jp}
\ba
\frac{S}{E} \leq \frac{2 \pi R}{\hbar}.
\label{eq:B-bound}
\ea
Although the presence of the black hole has been crucial for deriving \eqref{eq:B-bound}, the latter does not depend on any black hole parameter. In particular, there is no explicit dependence on $G$. It is stated that the Bekenstein bound is a non-gravitational statement. 
It has further been pointed out that satisfying the GSL may not be necessarily related to the existence of the bound \cite{Unruh:1983ir,Bekenstein:1999bh}.

A gravitational entropy bound explicitly depending on $G$ was proposed later and basically led to the idea of holography \cite{tHooft:1993dmi,Susskind:1994vu}. It states that the entropy in a spatial region enclosed by a sphere with area $A$ satisfies
\ba
S \leq \frac{A}{4 G \hbar}.
\label{eq:holo-bound}
\ea
The universal bound \eqref{eq:B-bound} applies to the ratio $S/E$. It is clear that a similar expression cannot be derived directly from the holographic bound \eqref{eq:holo-bound}.\footnote{The holographic entropy bound follows from the Bekenstein bound, if one assumes a weakly gravitating system, i.e. $r_\text{g} \lesssim R$. It has been argued that the former would validate the GSL \cite{Susskind:1994vu}, however, a counter-argument has been pointed out in \cite{Wald:1999vt}.} 

Since $E$ and $R$ do not necessarily depend on each other, it is not clear how to define these quantities separately when quantum wave functions are considered. Nevertheless, a rigorous derivation of \eqref{eq:B-bound} in a local quantum field theory (QFT) has been worked out by considering the vacuum subtracted entropy of arbitrary states defined in the Rindler region and re-expressing the product $ER$ as the modular energy \cite{Casini:2008cr}.
On the other hand, the holographic bound \eqref{eq:holo-bound} may be violated near singularities \cite{Wald:1999vt}. In order to avoid such difficulties, a covariant entropy bound that holds in a general curved spacetime has been proposed. Let $N$ be an outgoing future directed null hypersurface, where the expansion everywhere on the congruence $N$ shall be non-positive, $\theta \leq 0$, and no caustics on $N$ shall be allowed. If $B$ is a spacelike cross section of $N$ having area $A_B$, the covariant Bousso bound states, that
the future directed entropy flux $S$ through $N$, starting from $B$, is bounded as in \eqref{eq:holo-bound} with $A = A_B$ \cite{Bousso:1999xy}.
The generalized Bousso bound is the statement, that 
$A = A_B - A_{B'}$ in \eqref{eq:holo-bound}, where $B'$ denotes a second spacelike cross section in the future of $B$ at which $N$ terminates \cite{Flanagan:1999jp}.\footnote{The generalized Bousso bound implies a (tighter) version of the Bekenstein bound for a weakly gravitating setup \cite{Bousso:2002bh}.}

Except in the case of \eqref{eq:B-bound}, the gravitational entropy bounds are expressed in terms of the area $A$ of some codimension two surface divided by $4G\hbar$. Nevertheless, if the radius $R$ of the system with energy $M$ is taken to be the gravitational radius, i.e.
\ba
r_\text{g} = 2 G M,
\label{eq:SchRad}
\ea
then the Bekenstein bound can be re-expressed in form of a gravitational entropy bound as in \eqref{eq:holo-bound}, where $A = A_\text{h} = 4 \pi r_\text{g}^2$.
This is nothing but the earlier proposed Bekenstein-Hawking entropy \eqref{eq:S_BH}.
Indeed, the fact, that the Bekenstein bound gets saturated for the Schwarzschild black hole and may also hold for self-gravitating spherically symmetric material bodies \cite{Sorkin:1981wd}, indicates that the bound \eqref{eq:B-bound} would even universally apply in a (stable) gravitational spacetime, at least so long
\ba
R \geq r_\text{g}.
\label{eq:Rapprox}
\ea
This regime of validity implicitly assumes the presence of a (static) black hole in its Hartle-Hawking state. Thus, defining the system energy $E$ and size $R$ will not be problematic from the perspective of an outside observer. This may simply be done with respect to the thermodynamic sphere, i.e. $r=r_\text{g}$.
As mentioned above, the validity of the GSL may not depend on the universal validity of \eqref{eq:B-bound}. 
In this sense, there seems to be no particular argument why the latter should only be restricted to the semiclassical regime depicted in Fig.~\ref{fig:Sgen}. 

\section{Horizons, thermodynamics, (anti-)focusing}
\label{sec:htqf}

\begin{figure}[h]
  \centering
    \includegraphics[width=0.18\textwidth]{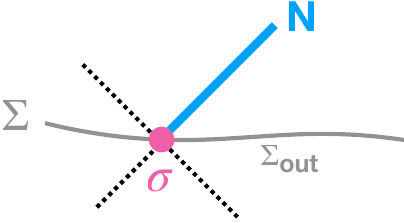}
    \caption{Codimension two surface $\sigma$ (purple dot) splits the Cauchy surface $\Sigma$ (gray curve) into two parts with $\Sigma_\text{out}$ lying in the exterior of $\sigma$. An orthogonal outgoing future directed null hypersurface $N$ (thick, blue line) is shot out from $\sigma$. $N$ does not need to be foliated by leaves.}
    \label{fig:congruence}
\end{figure}

Let $\sigma$ be a codimension two surface in a $3+1$ dimensional 
gravitational spacetime. The latter shall split a Cauchy surface $\Sigma$ into two parts, say $\Sigma_\text{in}$ and $\Sigma_\text{out}$. Let us further erect an orthogonal, outgoing future directed null hypersurface $N$ emanating from $\sigma$, where $N$ shall be divided into pencils of width of a small area element $\mathcal{A}$ around the null generators on $N$, see Fig.~\ref{fig:congruence}.
Next, introduce the expansion $\theta$ as the logarithmic derivative of $\mathcal{A}$.
In general relativity, the evolution of the congruence of null geodesics forming $N$ is described by the Raychaudhuri equation 
\ba
\theta' = - \frac{\theta^2}{2} - \zeta^2 - 8 \pi G T_{ab} k^a k^b,
\label{eq:Ray}
\ea
where the prime denotes a derivative with respect to the affine parameter along $N$ and $\zeta^2$ is the squared shear tensor. Here, we have used the equivalence between the null curvature condition and the NEC, $T_{ab} k^a k^b \geq 0$, which follows from the classical field equations, $G_{\mu \nu} = 8 \pi T_{\mu \nu}$, where $G_{\mu \nu}$ is the Einstein tensor and $T_{\mu \nu}$ denotes the stress energy tensor. Assuming that the NEC holds, implies the classical focusing theorem, $\theta' \leq 0$. 
Consider semiclassical gravity, $G_{\mu \nu} = 8 \pi \langle T_{\mu \nu} \rangle$.
The classical focusing theorem can be violated due to quantum fluctuations. Namely, since the NEC may not hold, it may follow from \eqref{eq:Ray} that light rays anti-focus, $\theta' > 0$. 
We consider the case, where the dividing surface $\sigma$ shall correspond to the horizon of a dynamical black hole, see Fig.~\ref{fig:evapBHsemi}. In other words, the former is taken to be the outermost marginally trapped surface \cite{Booth:2005qc}, $\theta = 0$, being equivalent to the apparent horizon. The gravitational piece in \eqref{eq:S_gen} shall be associated with the area of $\sigma$. The second contribution $S_\text{out}$ is taken to be the von Neumann entropy outside.
In analogy to the classical situation above, one may consider $4 G \hbar S_\text{gen}$ as some quantum corrected surface area. 
Assume that $\sigma$ is deformed along a single orthogonal generator on $N$. A quantum expansion, $\Theta$, can be defined as the derivative of $S_\text{gen}$ with respect to some affine parameter along $N$,
\ba
\Theta = \theta + \frac{4 G \hbar}{\mathcal{A}} S'_\text{out}.
\label{eq:Qexp}
\ea
The quantum focusing conjecture (QFC) is the statement that $\Theta' \leq 0$ \cite{Bousso:2015mna,Bousso:2015wca}.
For instance, let $\sigma$ be the black hole event horizon. One may initially assume quantum anti-focusing, $\Theta \leq 0$, which may apply to an old black hole. Using the QFC, the slice divided by $\sigma$ can be evolved along $N$ to a later slice divided by $\sigma'$. Since $S_\text{gen}[\sigma'] \leq S_\text{gen}[\sigma]$, this yields a quantum generalized Bousso bound \cite{Bousso:2014sda,Bousso:2014uxa,Bousso:2015mna}. 
In terms of $\Theta$, the statement of the GSL reads $\Theta \geq 0$.

\section{Unitarity and energy}
\label{sec:uec}

Consider now a spherical black hole in $3+1$ dimensional (asymptotically flat) spacetime. The spherically symmetric sector of $3+1$ dimensional Einstein gravity is described by the action of $1+1$ dimensional dilaton theory coupled to some quantum matter \cite{Grumiller:2002nm}. This description also permits the higher dimensional black hole solution. The dilaton in the dimensionally reduced theory plays the role of the spherical area. In the semiclassical regime, i.e. when the GSL holds, one may assume that the matter does not depend on the dilaton. In that case, one may work with the linearized Raychaudhuri equation \cite{Wall:2011kb}. Nevertheless, here we do not intend to solve some nontrivial near horizon quantum dynamics. 
So, we do not employ the mentioned linearization. We simply assume that the higher dimensional Schwarzschild black hole radiates massless particles. We may further assume that the emitted field modes have zero angular momentum (S waves), and effectively become spherically symmetric, so that they may be described by a $1+1$ dimensional CFT.\footnote{However, the arguments below may even apply when the field theory under consideration is not conformal invariant.}

Recall, that the gravitational radius $r_\text{g}$ shall determine the location of the dividing surface $\sigma$ being equivalent to the apparent horizon of the radiating black hole, i.e. $\theta^2=0$. We further remark that the squared shear tensor, $\zeta^2$, can be set to zero due to the effective $1+1$ dimensionality which also yields $\mathcal{A} = 1$ as the small area element in \eqref{eq:Qexp} becomes a point. Taking these aspects into account, we start from \eqref{eq:Ray} and use the quantum expansion in \eqref{eq:Qexp} to arrive at the general relation
\ba
8 \pi G \langle T_{vv} \rangle = 4 G \hbar  S''_\text{rad} - \Theta',
\label{eq:Tuu-1p1}
\ea
where $S_\text{rad}$, previously denoted as $S_\text{out}$, now corresponds to the entropy of the emitted radiation escaping into the future null infinity, $\mathscr{I}^+$. 
Recall that the expectation value of the matter stress energy tensor, $\langle T_{vv} \rangle$, in \eqref{eq:Tuu-1p1} is defined along the outgoing future directed null hypersurface $N$ emanating from the outermost marginally trapped surface $\sigma$.

Taking \eqref{eq:Tuu-1p1} and assuming the QFC leads to the known quantum null energy condition (QNEC), $2 \pi \langle T_{vv} \rangle \geq \hbar S''_\text{rad}$ \cite{Bousso:2015mna,Bousso:2015wca}. In the classical limit, the QNEC reduces to the classical focusing theorem, $\theta' \leq 0$. 

As it will be clear after certain reformulations, the quantum focusing regime, $\Theta' \leq 0$, basically captures the time window around $u_\text{P}$ for an evaporating black hole whose radiation entropy evolves unitarily. The quantum anti-focusing regime, $\Theta' \geq 0$, applies to the stages much before and after $u_\text{P}$.

In order to arrive at this picture, first recall that a quantum black hole can be considered as a fast scrambler \cite{Sekino:2008he}. Based on this feature, the time evolution of its radiation entropy can be computed \cite{Page:2013dx}. Taking such properties into account, as well as the discussion above, we may write\footnote{The exact evolution around the (retarded) time $u = u_\text{P}$ would, of course, not be identical to the simplified approach in \cite{Page:2013dx} developing a kink around $u = u_\text{P}$. We may instead assume that the entropy curve is sufficiently smooth around the Page time when computed within the hypothetical unitary fine grained description.}
\ba
\Theta' \simeq 8 G \hbar \times
 \begin{cases}
    (\Omega/2) S''_\text{rad} & \text{for $u \lesssim u_\text{P}$}\\
    S''_\text{rad} & \text{for $u \gtrsim u_\text{P}$}
  \end{cases}
  \label{eq:Theta-p-absch}
\ea
for an evaporating, spherical quantum black hole, where $0 \leq \Omega$.
For instance, a $3+1$ dimensional black hole unitarily emitting massless particles into empty space yields $\Omega \approx 1/3$ \cite{Page:2013dx}.\footnote{In a realistic unitary evaporation process, the constant $\Omega$ appearing in \eqref{eq:Theta-p-absch} may generally be non-zero as the increasing rate of $S_\text{rad}$ would exceed the decreasing rate of $S_\text{BH}$ for $u < u_\text{P}$, i.e. $S'_\text{out} \equiv S'_\text{rad} > - S'_\text{BH}$ \cite{Page:2013dx}, where $\Theta \simeq S'_\text{gen}$ by definition. However, without loss of generality, we may consider the case $\Omega = 0$ for simplifying reasons. In other words, we may assume the saturation of the QNEC.}

In the case of a symmetric Schwarzschild black hole formed from a collapsing null shell, whose radiation modes, i.e. the Bogoliubov coefficients, as seen from the far distance (i.e.~in $\mathscr{I}^+$), are effectively described by a $1+1$ dimensional conformal theory, we may in fact write \cite{Akal:2021foz}
\ba
8 \pi G \langle T_{uu} \rangle = 4 G \hbar  S''_\text{rad}.
\label{eq:Tuu-1p1-v2}
\ea
Here, the numerical values of the constants on both sides of \eqref{eq:Tuu-1p1-v2} are not of particular relevance for the present discussion. Importantly, however, the prime appearing in \eqref{eq:Tuu-1p1-v2} (and in what follows) now denotes the derivative with respect to the affine parameter $u$ defined along future null infinity, $\mathscr{I}^+$.
Similarly, the expectation value of the energy momentum tensor in \eqref{eq:Tuu-1p1-v2}, $\langle T_{uu} \rangle$, is defined in $\mathscr{I}^+$.
We may absorb the numerical prefactors in \eqref{eq:Tuu-1p1} and \eqref{eq:Theta-p-absch} into a $u$-dependent differentiable, smooth function $\hbar \mathcal{F}(u) \rightarrow \mathcal{F}(u)$. This function regulates the behavior of the prefactor in the quantum expansion \eqref{eq:Theta-p-absch} which, as we remember, has followed from simple unitarity considerations. Together with \eqref{eq:Tuu-1p1-v2}, we may finally write
\ba
\langle T_{uu} \rangle =  \mathcal{F}(u) S''_\text{rad}.
\label{eq:Tuu-1p1-v3}
\ea

Since it is assumed that the black hole is formed from a pure state, we may continue with the following unitarity condition\footnote{As pointed out before, we here assume that the evolution of $S_\text{rad}$ is fully captured by a smooth curve, at least, during the pre-Planckian era of the unitary radiation process.} 
\ba
S_\text{rad}(\pm \infty) = 0 = S'_\text{rad}(u_\text{P}),
\label{eq:UC}
\ea
where $S_\text{rad}(u_\text{P}) = \mathrm{max}_u \{ S_\text{rad}(u) \}$.
The total energy during the evaporation process should be finite and the averaged NEC (ANEC) satisfied along $\mathscr{I}^+$, i.e.
$0 \leq \int_{\mathscr{I}^+} du\ \langle T_{uu} \rangle < \infty$.\footnote{In general, the (semiclassical) NEC can be violated by quantum
effects, even in flat spacetime, which happens in a finite time window. However, the ANEC may still be valid in
QFT despite the mentioned temporal violation of NEC. This is precisely what is assumed to be happening in
the present discussion. The NEC is violated along the future null infinity, $\mathscr{I}^+$, namely, temporally around
the Page time $u_\text{P}$, but $\langle T_{uu} \rangle $ integrated along the entire $\mathscr{I}^+$ leads to the validity of the ANEC.

On the other hand, the reader might think that the violation of the NEC would imply that unphysical radiation escapes into the future null infinity. However, there will be nothing strange from the perspective of an outside observer. The finite amount of radiation collected in a finite time window will be thermal and the mentioned unitarity constraints do not apply. The observer would therefore find the usual thermal Hawking quanta collected in a finite time interval. Unitarity/energy consistency is only restored if
the entire $\mathscr{I}^+$ physics is taken into account. Eventually, this is what the radiation itself will
experience. It is not appropriate to consider a finite time window
and, thus, a fraction of all radiated quanta in order to make the situation consistent with the preservation of information.}

Thus, by making use of \eqref{eq:Tuu-1p1-v3} and imposing unitarity in the form of the Page curve, i.e. \eqref{eq:UC}, it can be shown that 
\ba
S_\text{rad}' (\pm \infty) = 0. 
\label{eq:ANECC}
\ea
By further employing the conditions \eqref{eq:UC}, \eqref{eq:ANECC} and rewriting $S''\exp(S') = (\exp(S'))'$, we end up with definite integrals of the form
\ba
 \int_{-\infty,u_\text{P}}^{u_\text{P},+\infty} du\ \left(\exp(S'_\text{rad})\right)'.
\ea
Finally, bringing all contributions together yields
\ba
\int_{\mathscr{I}^+} du\ \langle T_{uu} \rangle \exp \left(  S'_\text{rad} \right) = 0.
\label{eq:clos-int}
\ea
As $\exp(S'_\text{rad}) > 0$ for $u=[-\infty,+\infty]$, the result \eqref{eq:clos-int} implies that $\langle T_{uu} \rangle$ violates the NEC in a finite time window centered around some $\tilde u$. It can be shown that the minimum, i.e. $\langle T_{uu} \rangle' = 0$, is located at $u = u_\text{P}$, means $\tilde u = u_\text{P}$. Let us consider a young black hole, i.e. $u < u_\text{P}$. Suppose that its radiation entropy $S_\text{rad}$ follows a symmetric curve, i.e. $\Omega = 0$, then \eqref{eq:Tuu-1p1-v2} reduces to the saturated QNEC, $2 \pi \langle T_{uu} \rangle = \hbar S''_\text{rad}$.\footnote{Symmetric curve here means $S'_\text{BH} = - S'_\text{rad}$ for $u < u_\text{P}$. Regarding the saturation of the QNEC in holography, refer, for instance, to \cite{Leichenauer:2018obf}. However, note that in the present scenario, deduced from the \textit{relocated} relation \eqref{eq:Tuu-1p1-v2}, the saturation instead applies for the expectation value of the energy momentum tensor, $\langle T_{uu} \rangle$, defined in $\mathscr{I}^+$.}

Notably, in the case of $1+1$ dimensional CFT, one may consider the particular combination
\ba
\tilde{S} \equiv S_\text{rad}'' +  (6/c) (S_\text{rad}')^2, 
\label{eq:Stilde}
\ea
where the parameter $c$ denotes the corresponding central charge. The appearance of the second term in the definition \eqref{eq:Stilde} goes back to the underlying symmetry \cite{Wall:2011kb}. The saturated QNEC then becomes of the form $2 \pi \langle T_{uu} \rangle = \hbar \tilde S$. In this case, if the function \eqref{eq:Stilde} is re-expressed as $\tilde S = \exp(-\frac{6}{c}S_\text{rad}) [ \exp(\frac{6}{c}S_\text{rad}) ]''$, the negativity of $\langle T_{uu} \rangle$, as we have obtained above, follows from an analogous definite integral of the form $\int_{\mathscr{I}^+} du\ \langle T_{uu} \rangle \exp \left(  \frac{6}{c}S_\text{rad} \right) = 0$.
 
The (modified) QNEC introduced above is, for instance, saturated for the radiation entropy holographically computed in a two dimensional moving mirror CFT modeling the formation and evaporation of a black hole including backreaction \cite{Akal:2020twv}. In fact, such a boundary field theory can be interpreted as a two dimensional dilaton gravity coupled to conformal matter \cite{Akal:2021foz}. This understanding thus aligns with the general considerations presented here.

We shall note that we here do not rely on any specific boundary field theory setup of the mentioned type nor do we make specific assumptions about the near horizon aspects of the black hole. We only assume the validity of unitarity (in the form of the Page curve) as seen by a far distant observer sitting in $\mathscr{I}^+$. Furthermore, the general assumptions underlying the present derivations are independent of the exact shape of the fine grained radiation entropy curve. We stick to the most general assumptions which turn out to be sufficiently justified for a fine grained entropy curve reflecting information preservation.

\section{Information and quantum entropy sphere}
\label{sec:ises}

Let us reflect on what we have arrived at so far. We have considered a spherically symmetric black hole radiating into empty space. We have required that the radiation escaping to $\mathscr{I}^+$ encodes the right amount and structure of quantum information so that unitarity is maintained. This has made $S_\text{rad}$ follow the Page curve. In the following, we elaborate on how the findings above reconcile with the laws of black hole thermodynamics, particularly, in light of the discussed universal entropy bounds.
\begin{figure}[h]
  \centering
    \includegraphics[width=0.18\textwidth]{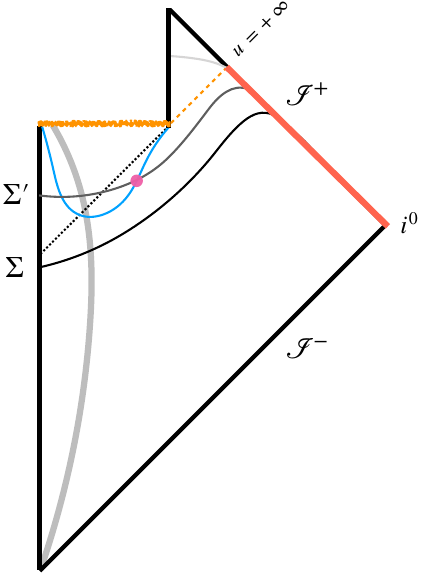}
    \caption{Global, semiclassical spacetime of an evaporating black hole formed by gravitational collapse. The outermost marginally trapped surface (blue curve), i.e. apparent horizon with $\theta=0$, defining the thermodynamic sphere with radius $r=r_\text{g}$, lies outside the event horizon (dotted line). Emitted Hawking particles escape into the future null infinity, $\mathscr{I}^+$. Seen from the point, where the latter crosses the Cauchy horizon (dashed, yellow line), the black hole has completely evaporated, i.e. $M = 0$.}
    \label{fig:evapBHsemi}
\end{figure}  
The mass of the black hole is taken to be the usual ADM mass defined in $i^0$ shown in Fig.~\ref{fig:evapBHsemi}. Let $M_0$ be the initial mass of the black hole formed from a pure state.
According to what we have obtained, the mass evolution seen on $\mathscr{I}^+$ at $u=u_\text{P}$ is given by
$E_{+}(u_\text{P}) = M_0 + \int_{-\infty}^{u_p} du\ \dot{E}_+$,
where $\dot{E}_+ \equiv - \langle T_{uu} \rangle$.
Semiclassically, the Schwarzschild black hole is characterized by its mass $M$. The horizon surface area $A_\text{h}$ including all thermodynamic quantities, and thus the rate of energy loss, are a function of $M$.
Seen from the outside, the black hole emits almost perfect blackbody radiation with Hawking temperature $T_\text{H} = \hbar r_\text{g}/A_\text{h}$. Assuming that only massless particles are emitted and the (apparent) horizon at $r=r_\text{g}$ is the surface responsible for radiation, but not necessarily the place where it is generated, the Stefan-Boltzmann law together with $T_\text{H}$ leads to the standard thermodynamic rate of mass loss
$\dot M \simeq - \hbar/(\pi G^2 M^2)$. If the initial mass $M_0$ is taken to be sufficiently large, then the black hole will still be large enough at $u_\text{P}$. Thus, energy conservation should agree with the mass decrease given by 
$M(u_\text{P}) = M_0 + \int_{-\infty}^{u_p} du\ \dot M$.
Demanding $M = 0$ and using $\dot M$, one arrives at the usual evaporation time, $u_\text{evap} \simeq \pi G^2 M_0^3/\hbar$. 

Let us now scrutinize the derivations in the previous paragraph. We have seen that $\dot{E}_+(u_\text{P}) > 0$, where $\mathrm{min}_u \{ -\dot{E}_+ \} = u_\text{P}$, therefore 
\ba
E_{+}(u_\text{P})> M(u_\text{P}). 
\label{eq:estimate}
\ea
To make the estimation \eqref{eq:estimate} more evident, we first start with an implication of the unitarity condition in \eqref{eq:UC}. Namely, according to the latter, we may write 
\begin{equation}
    S_\text{rad}(u_\text{P}) < S_\text{therm} (u_\text{P}),
\end{equation}
where $S_\text{therm}$ shall denote the monotonically increasing radiation entropy in the standard, unitarity violating semiclassical description.
Since, unitarity in the form of the Page curve implies $S_\text{rad}(u_\text{P}) = \mathrm{max}_u\{S_\text{rad}(u)\}$, we may conclude
\begin{equation}
    S'_\text{rad}(u_\text{P}) < S'_\text{therm}(u_\text{P}),
\end{equation}
or, equivalently, 
\begin{equation}
    -S'_\text{rad}(u_\text{P}) > -S'_\text{therm}(u_\text{P}).
    \label{eq:Sestimate}
\end{equation}
For the definite evolution integrals, we, thus, straightforwardly get
\ba
\begin{split}
    M_0 + \int^{u_\text{P}}_{-\infty} du\ \dot{M} &= M(u_\text{P})\\
    M_0 + \int_{-\infty}^{u_\text{P}} du\ \dot{E}_+ &= E_+(u_\text{P})
    \end{split}
\ea
with $M(-\infty) = E_+(-\infty) = M_0$. Here, we may think of some collapsing null shell forming the black hole with initial mass $M_0$. 
Furthermore, by definition we have
\begin{equation}
    \dot{E}_+(u) = - \langle T_{uu} \rangle = - S''_\text{rad}(u),
\end{equation}
so that
\begin{equation}
    E_+(u) = - S'_\text{rad}(u)
    \label{eq:Eplus-Sminus}
\end{equation}
Accordingly, using the estimation \eqref{eq:Sestimate} and the relation \eqref{eq:Eplus-Sminus} gives rise to the inequality \eqref{eq:estimate}, where $-S'_\text{therm}(u_\text{P}) \equiv M(u_\text{P})$.

Now, for a moment, consider $r_+ = 2 G E_+$ to be the gravitational radius. As initially discussed, the ratio $S/E$ may be bounded from above as \eqref{eq:B-bound}. Assume, for instance, that the latter is saturated, as it is so for the Schwarzschild black hole, means $S/E_+ = 2 \pi r_+/\hbar$, where $S/E_+ > S_\text{BH}/M = 2 \pi r_\text{g}/\hbar$.
However, this is not consistent with what we have required in order to arrive at the current situation. Namely, for $u \geq u_\text{P}$, unitarity in the sense discussed above implies
\ba
S = S_\text{rad}(u) = S_\text{BH}(u)
\label{eq:cond}
\ea
and, particularly, $dS_\text{rad} \leq 0$ so that $dS_\text{BH} > 0$ cannot be allowed, which would, therefore, violate the GSL \eqref{eq:GSL}.
If we respect unitarity and take into account that the laws of black hole mechanics hold at this stage, it follows that $S_\text{BH}/E_+ \leq 2 \pi r_+/\hbar$.
Rewriting this inequality as $S/M \leq 2 \pi (E_+/M)r_+/\hbar$, we end up with
\ba
\frac{S_\text{BH}(u_\text{P})}{M(u_\text{P})} \leq \frac{2 \pi r_\text{g}}{\hbar}
\left[ 1 + \frac{\Delta_+}{M(u_\text{P})} \right]^2,
\ea
where $E_+(u_\text{P}) \equiv M(u_\text{P})  + \Delta_+$. The second term $\Delta_+$ shall denote some positive correction term.
This is the (extended) quantum entropy sphere bounding the ratio $S/E$ for the unitarily evaporating black hole. Its radius, $R_\text{qs}$, exceeds the gravitational radius defining the thermodynamic sphere. 
Extrapolating to the initial stage of the black hole having mass $M_0$, we may therefore write
\ba
\frac{{S_\text{BH}}_0 }{M_0} \leq \frac{2 \pi R_\text{qs}}{\hbar}.
\ea
We conclude that the number of (quantum) bits associated with the initial coarse grained entropy ${S_\text{BH}}_0$ will be contained in a quantum entropy sphere of radius $R_\text{qs}$ satisfying
\ba
R_\text{qs} > r_\text{g}.
\ea 
In what follows, we summarize the findings and conclude with some final remarks on the black hole singularity.

\section{Discussion}
\label{sec:}

Let be given a black hole of finite mass formed by gravitational collapse, with an entropy that is determined to the leading order by the Bekenstein-Hawking formula. If the total energy is conserved and the fine grained radiation entropy follows a unitary entropy curve, we have shown that the maximal amount of information, required to fully quantum mechanically describe the black hole system, has to be contained in a quantum entropy sphere whose radius $R_\text{qs}$ exceeds the gravitational radius $r_\text{g}$. The latter is determined by the outermost marginally trapped surface being equivalent to the apparent horizon. Such an enlargement is understood to be the consequence of purely quantum gravitational effects implying that radiation gets purified via semiclassically invisible quantum correlations extending across the black hole atmosphere \cite{Akal:2020ujg}. 

If a trapped surface, satisfying the laws of black hole mechanics \cite{Hayward:1993mw}, forms and the NEC is fulfilled, a gravitational singularity will inevitably exist in the future of that surface. This is the well known singularity theorem \cite{Penrose:1964wq}. In the semiclassical regime, i.e. when the GSL holds, the generalized entropy can be used to define a quantum corrected version of the trapped surface. It turns out that a singularity in the future of such a surface still persists \cite{Wall:2010cj,Wall:2010jtc}. 

When the fine grained radiation entropy enters its decreasing phase, the GSL becomes invalidated and the semiclassical description breaks down. This is argued to be happening due to the gradual transfer of the near horizon information purifying the Hawking radiation.
The usual arguments regarding the notion of a future singularity will therefore not be applicable. In fact, the described quantum gravitational effects cannot be made compatible with a global, semiclassical spacetime \cite{Akal:2020ujg} and thus invalidate the notion of a classical singularity within the deep interior of the black hole, namely, long before the Planckian final stage. 
Indeed, it is to be expected that, in order to resolve the known tensions arising within the semiclassical description, the purification mechanism and the invalidation of the classical singularity will not act mutually independently. Instead, as put forward in the present work, their resolutions will be closely interconnected.

\section*{Acknowledgement}
I am grateful to Tadashi Takayanagi for the valuable comments. I further thank Francesco Di Filippo, Mohammad Ali Gorji, Tatsuma Nishioka, Yasunori Nomura, Tadashi Takayanagi, Tomonori Ugajin, and Zixia Wei for the informative discussions.
I acknowledge the support from the Japan Society for the Promotion of Science and the Alexander von Humboldt foundation. I am supported by a JSPS Grant-in-Aid for Scientific Research under the Grant No.~19F19813.

\bibliographystyle{JHEP}
\bibliography{NHQCBB_bib}

\end{document}